\documentclass[aps,pre,twocolumn,superscriptaddress,showpacs,floatfix]{revtex4}

\usepackage{amsmath,amssymb}
\usepackage[dvips]{graphicx}

\begin{document}

\title{Morphological stability of electromigration-driven vacancy
islands}

\author{Frank Hau{\ss}er}
\email{hausser@caesar.de}
\affiliation{Crystal Growth Group, Research Center caesar,
    Ludwig-Erhard-Allee 2,
    53175 Bonn, Germany}
\author{Philipp Kuhn}
\email{philipp@thp.uni-koeln.de}
\affiliation{Institut f\"ur Theoretische Physik, Universit\"at zu K\"oln,
Z\"ulpicher Strasse 77, 50937 K\"oln, Germany}
\author{Joachim Krug}
\email{krug@thp.uni-koeln.de}
\affiliation{Institut f\"ur Theoretische Physik, Universit\"at zu K\"oln,
Z\"ulpicher Strasse 77, 50937 K\"oln, Germany}
\author{Axel Voigt}
\email{voigt@caesar.de}
\affiliation{Crystal Growth Group, Research Center caesar,
    Ludwig-Erhard-Allee 2,
    53175 Bonn, Germany}
\affiliation{Institut f\"ur Wissenschaftliches Rechnen, Technische
  Universit\"at Dresden, Zellescher Weg 12-14, 01062 Dresden}
\date{\today}

\begin{abstract}
  The electromigration-induced shape evolution of two-dimensional vacancy
  islands on a crystal surface is studied using a continuum approach. We
  consider the regime where mass transport is restricted to terrace diffusion
  in the interior of the island. In the limit of fast attachment/detachment
  kinetics a circle translating at constant velocity is a stationary solution
  of the problem. In contrast to earlier work [O. Pierre-Louis and T.L.
  Einstein, Phys. Rev. B \textbf{62}, 13697 (2000)] we show that the circular
  solution remains linearly stable for arbitrarily large driving forces. The
  numerical solution of the full nonlinear problem nevertheless reveals a
  fingering instability at the trailing end of the island, which develops from
  finite amplitude perturbations and eventually leads to pinch-off.  Relaxing
  the condition of instantaneous attachment/detachment kinetics, we obtain
  non-circular elongated stationary shapes in an analytic approximation which
  compares favorably to the full numerical solution.
\end{abstract}





\pacs{05.45.-a, 68.65.-k, 66.30.Qa, 68.35.Fx}

\maketitle

\section{Introduction}
\label{s:introduction}

Much of the diversity of natural shapes in the inanimate world is the result
of morphological instabilities.  The paradigmatic example is the
Mullins-Sekerka instability, in which a spherical solid nucleus in an
undercooled melt forms lobes and petals which eventually develop into a
delicate dendritic pattern \cite{Pelce2004}, and its two-dimensional analogs
that can be observed in the growth of islands on crystal surfaces
\cite{Michely2004}.  These systems share the common mathematical structure of
moving boundary value problems, in which an interface evolves in response to
the gradient of some continuous field defined in the spatial domains that it
separates. In the context of two-dimensional crystal surfaces, the interfaces
are atomic height steps separating different terraces, and their motion is
governed by the attachment and detachment of the adsorbed atoms (adatoms)
\cite{Jeong1999,PierreLouis2005,Krug2005}.

A rich variety of two-dimensional morphological instabilities has been
observed on the surfaces of current-carrying crystals, where an
electromigration force induces a directed motion of adatoms
\cite{Yagi2001,Minoda2003}.  The microscopic origin of this force is a
combination of momentum transfer from the conduction electrons (the 'wind
force') and a direct effect of the local electric field \cite{Sorbello1998}.
On stepped surfaces vicinal to Si(111), electromigration has been found to
cause step bunching \cite{Latyshev1989}, step meandering \cite{Degawa1999},
step bending \cite{Thuermer1999} and step pairing \cite{PierreLouis2004}
instabilities.  In addition, single layer adatom islands on silicon surfaces
have been seen to drift under the influence of electromigration
\cite{Metois1999,Saul2002}.

In the present paper we focus on the morphological stability of single layer
vacancy islands driven by electromigration. We build on the work of
Pierre-Louis and Einstein (PLE) \cite{PierreLouisEinstein2000}, who introduced
a class of continuum models for island electromigration. The different models
are distinguished according to the dominant mechanism of mass transport, which
can be due to periphery diffusion (PD) along the edge of the island, terrace
diffusion (TD), or two-dimensional evaporation-condensation (EC), i.e.
attachment-detachment, kinetics. In the PD regime the dynamics of the island
edge is local, while in the TD and EC regimes it is coupled to the adatom
concentration on the terrace. In the TD (EC) regime diffusion is slow (fast)
compared to the attachment/detachment processes, as reflected in the magnitude
of the kinetic lengths 
\begin{equation}
\label{kinlengths}
d_\pm = D/k_\pm
\end{equation}
defined as the ratio of the surface diffusion constant $D$ to the rates
of adatom attachment to a step from the lower ($k_+$) or upper ($k_-$) terrace,
respectively. The two rates generally differ, because step edge barriers
suppressing attachment across descending steps are ubiquitous on many surfaces,
leading to $k_+ > k_-$ \cite{Michely2004}.  

As a common feature of the models of PLE, the electromigration force is taken
to be of constant direction and magnitude everywhere, which implies in
particular that it is not affected by the presence and the shape of the
island. This is motivated by the fact that the island constitutes a small
perturbation in the morphology of the crystal, which is not expected to
substantially change the distribution of the electrical current in the bulk.
Detailed atomistic calculations do in fact show that the electromigration
force is modified in the vicinity of a step \cite{Rous1999,Rous2000}, which
may also be incorporated into a continuum model
\cite{PierreLouis2006}, but this
is a higher order effect that can be neglected on the present level of
description. The situation is completely different for electromigration-driven
macroscopic voids in metallic thin films, which can be modeled using a closely
related two-dimensional continuum approach
\cite{Ho1970,Wang1996,Mahadevan1996,Schimschak1998,Gungor1999,Schimschak2000,Cummings2001}.
Since the void interrupts the current flow, the effect of the void shape on
the current distribution is an essential part of the analysis. In the
following we refer to this problem as \textit{void migration}, to be
distinguished from \textit{island migration} under a constant force.

Apart from the work of PLE, analytical results concerning the morphological
stability of electromigration-driven two-dimensional shapes have been obtained
only in the PD regime.  In the absence of crystal anisotropy the basic
solution is then a circle moving at constant velocity \cite{Ho1970}. In the
case of island electromigration the circle becomes linearly unstable at a
critical radius or critical driving force \cite{Wang1996}. Beyond the linear
instability stationary shapes that are elongated in the current direction
appear \cite{Kuhn2005a,Kuhn2005b}. Remarkably, the stability scenario for void
migration is completely different.  Voids are linearly stable at any size
\cite{Wang1996,Mahadevan1996}, but they become nonlinearly unstable beyond a
finite threshold perturbation strength, which decreases with increasing radius
or driving force \cite{Schimschak1998,Schimschak2000}. Unstable voids break up
into smaller circular voids, and non-circular stationary shapes do not exist
\cite{Cummings2001}.  The increasing sensitivity to finite amplitude
perturbations can be linked to the increasing non-normality of the linear
eigenvalue problem, which leads to transient growth of linear perturbations
\cite{Schimschak1998}. This route to nonlinear instability has been previously
described for linearly stable hydrodynamic flows
\cite{Trefethen1993,Grossmann2000,Trefethen2005}, and it will be further
discussed below in Section~\ref{s:linStability}.

\begin{figure}[htp]
  \includegraphics*[width=0.8\columnwidth]{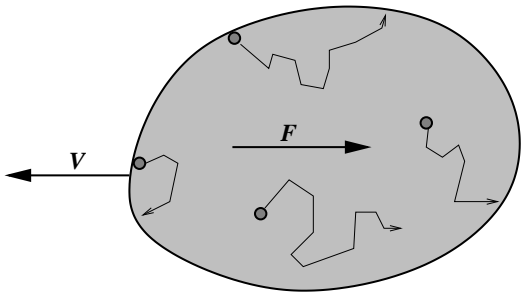}
  \caption{Schematic of the interior model. Adatoms detach from the inner boundary of the
vacancy island and diffuse subject to the electromigration force $F$ directed to the
right. As a consequence, the entire island drifts to the left at speed $V$.}
  \label{f:interior}
\end{figure}

In the present paper we focus on the ``interior model'' introduced by PLE.
Referring to Fig.~\ref{f:interior}, we consider a single vacancy island which
is isolated from the surrounding upper terrace by a strong step edge barrier
that prevents adatoms from entering across the descending step ($k_- = 0, \;
d_- = \infty$).  Diffusive motion of vacancy islands mediated by internal
terrace diffusion has been observed experimentally on the Ag(110) surface
\cite{Morgenstern2001}. The mathematically equivalent process of internal
diffusion of \textit{vacancies} also plays an important role in the motion of
\textit{adatom} islands \cite{Heinonen1999}. We will use the terminology
appropriate for adatom diffusion inside a vacancy island throughout the paper.

The mathematical description of the interior model leads to a moving boundary
value problem on a finite domain, which we formulate in the next section. A
key ingredient of our analytic work is a separation ansatz for the adatom
concentration, which allows us to determine stationary island shapes and
investigate their linear stability in a simpler and more transparent way than
in previous work \cite{PierreLouisEinstein2000}. The analytic approach is
complemented by numerical simulations of the full nonlinear and nonlocal
dynamics.  Specifically, in Section~\ref{s:steadyStates} we compute
non-circular stationary island shapes perturbatively in the parameter $\delta
= d_+/R_0$, where $R_0$ is the radius of the circle which solves the
stationary problem in the TD limit ($d_+ = 0$).  Section~\ref{s:linStability}
is devoted to the linear stability analysis of the circular solution for $d_+
= 0$. In agreement with PLE, we find that the eigenvalues of the linearized
problem depend only on the ratio $z = R_0/\xi$, where
\begin{equation}
\label{xi}
\xi = \frac{k_\mathrm{B} T}{F}
\end{equation}
is the characteristic length scale associated with the electromigration force
$F$. However, in contrast to PLE, who argued (on the basis of a less extensive
analysis) that the circle becomes unstable for $z > 0.1$, we show that it in
fact \textit{remains linearly stable for all values of $z$}. Simulations of
the full nonlinear evolution in Section~\ref{s:nonLinEv} nevertheless reveal
an instability under finite amplitude perturbations, in which a finger
develops at the trailing end of the island and eventually leads to a
pinch-off.  In Section~\ref{s:conclusions} we summarize our results and
discuss their significance in the broader context of morphological stability
in moving boundary value problems.
 
%
%
%
\section{Model}
\label{s:model}

Since we assume that the mass transport on the surface is dominated by
terrace diffusion, the main dynamical quantity of interest is the adatom concentration 
$c(x,y,t)$ on the terrace. By mass conservation its time evolution is governed by
\begin{align}
  \label{eq:diff}
  \partial_t  c + \nabla \cdot  \vec{j} = 0 \\
  \vec{j} = -D\nabla c + \frac{D}{\xi}\hat{x}  c,
\end{align}
where the current $\vec{j}$ takes into account the contributions from
diffusion and electromigration. Since we assume the force to be constant,
electromigration appears as a drift in the direction of the electric
current, denoted by the unit vector $\hat{x}$.

The coupling to the periphery of the island is given by the boundary
conditions at the step edge. Let the subscripts $+,\,-$ denote
quantities at the lower and upper terrace, respectively, $\vec{n}$ the
normal pointing from the upper to the lower terrace, and $v$
the normal velocity of the island boundary. The fluxes
\begin{equation}
\label{eq:fluxes}
j_\pm : = \mp (\vec{j_\pm} \cdot \vec{n} - c_\pm v)
\end{equation}
from the lower ($+$) and the upper ($-$) terrace, respectively, 
towards the step are then
assumed to be proportional to the deviation from equilibrium \cite{Krug2005}, i.e.,
\begin{align}
\label{eq:bc}
   j_\pm =  k_\pm ( c_\pm -  c_\mathrm{eq}),
\end{align}
with $k_+,\, k_-$ denoting the attachment rates from the lower and
upper terrace, respectively. Here the equilibrium density $c_\mathrm{eq}$ is
given by the linearized Gibbs-Thomson relation
\begin{equation}
  \label{eq:eqDensity}
   c_\mathrm{eq} = c^0_\mathrm{eq}(1 + \Gamma\kappa),
   \qquad \Gamma = a^2\tilde{\gamma} /k_\mathrm{B} T,
\end{equation}
where  $\tilde{\gamma}$ denotes the (isotropic) step stiffness, $a$ the lattice
constant, and $\kappa$ the curvature of the terrace boundary. 
The validity of (\ref{eq:eqDensity}) requires the capillary length
$\Gamma$ to be small compared to the radius of curvature
of the island boundary. Defining
the kinetic lengths by (\ref{kinlengths}), we note that in the
terrace diffusion limit (TD), where the attachment/detachment becomes
instantaneous ($k_\pm \to \infty$, $d_\pm \to 0$), the boundary
conditions \eqref{eq:bc} reduce to
\begin{equation}
  \label{eq:bcTD}
  c = c_\mathrm{eq}.
\end{equation}
Finally, by mass conservation, the normal velocity  $v$ of the boundary is
\begin{align}
\label{eq:vel}
  v = a^2(j_+ + j_- ).
\end{align}
We neglect periphery diffusion, since we are concerned here with
the kinetic regime where terrace diffusion is the dominant 
mass transport mechanism. For the interior model $k_- = 0$,
which implies $j_- = 0$, and (\ref{eq:bcTD}) applies on the interior
(lower) terrace in the TD limit.  

In the following, it is assumed that the electric current is in 
the $x$-direction,
i.e. $\hat{x} = (1,0)$. Moreover, for analytic calculations 
the quasistatic approximation is frequently used, 
which amounts to setting $\partial_t c = 0$ in 
the diffusion equation (\ref{eq:diff}), which then 
reduces to
\begin{equation}
  \label{eq:diffQs}
    \Delta c - \xi^{-1}\partial_x c = 0
\end{equation}
and omitting the term proportional to $v$ in (\ref{eq:fluxes}), which arises
from the sweeping of adatoms by the moving step.
For our purposes, the general solution of (\ref{eq:diffQs}) is most
conveniently expressed in polar coordinates. To arrive at a suitable
representation, we first eliminate the drift term breaking the
rotational symmetry of (\ref{eq:diffQs}) via the ansatz
\begin{align*}
c(x,y)= \exp(\frac{x}{2\xi})f(\frac{x}{2\xi},\frac{y}{2\xi}),
\end{align*}
which leads to the Helmholtz equation $\Delta f = f$. 
Separation of the latter equation yields a harmonic
angular dependence and a modified Bessel function of imaginary
argument $I_n$ for the radial part. The general solution of (\ref{eq:diffQs}) 
is then a superposition of the form 
\begin{align}
  \label{eq:sep1}
  c(r,\theta) &= \exp(\tfrac{r}{2\xi}\cos\theta) f(r,\theta),  \\
  \label{eq:sep2}
  f(r,\theta) &=
  \sum_{n=-\infty}^{\infty}\hat{c}_n I_n(\tfrac{r}{2\xi})\exp(in\theta),
\end{align}
where the unknown coefficients $\{ \hat{c}_n \}$ are to be determined by the
boundary conditions \eqref{eq:bc}.

We further note that, in the quasistatic approximation,
the total area $A$ of the island
is strictly conserved by the dynamics. For the interior model
one computes
\begin{align*}
\frac{d}{d t} A = \int_{\partial \Omega} v \;  ds &= -a^2
\int_{\partial \Omega} \vec{j}_+\cdot \vec{n} \;ds \\
 &= -a^2 \int_\Omega \nabla \cdot \vec{j}_+ \; dA = 0,
\end{align*}
using the divergence theorem, where 
$\Omega$ denotes the interior domain and $\partial \Omega$
its boundary. The last integral vanishes
in the quasistatic approximation. This means that the mass
exchange between the island boundary (the bulk)  and the adatom concentration
inside is always balanced. In this sense the diffusion field merely mediates the
mass transport from one part of the boundary to another.
%
%
%
%
\section{Steady states}
\label{s:steadyStates}
For the interior model in the TD limit ($d_+ = 0, \; d_- = \infty$), circular
islands are steady states. More precisely, an island with radius $R_0$, constant
adatom concentration
 \begin{equation*}
  c = c_0 = c^0_\mathrm{eq}(1 - \tfrac{\Gamma}{R_0})
 \end{equation*}
 and drifting with constant velocity
\begin{equation}
\label{eq:velDrift}
 \vec{V} = - \frac{D}{\xi}\frac{a^2 c_0}{1 - a^2 c_0} \hat{x}
\end{equation}
is a solution of \eqref{eq:diff}-\eqref{eq:vel}; 
the factor $(1 - a^2 c_0)^{-1}$ is a correction
to the quasistatic approximation, which requires $a^2 c_0 \ll 1$. However, if the
attachment is not instantaneous ($d_+ > 0$), the circle is no longer stationary.  As
was shown in \cite{PierreLouisEinstein2000}, an expansion of the
interior model to second order in $z = R_0/\xi$ leads to noncircular steady
states being elongated perpendicular to the field direction.

In the following we will investigate the existence of steady states in the
regime where $z \sim 1$. To this end, we expand the interior model in the small
parameter
\begin{equation*}
  \delta = d_+ / R_0
\end{equation*}
and look  for first order
perturbations of the steady state, i.e.
\begin{align*}
  R(\theta) &= R_0  + \rho(\theta) + {\mathcal O}(\delta^2), \\
  \qquad c(r,\theta) &= c_0 + c_1 + {\mathcal O}(\delta^2).
\end{align*}
Applying the quasistatic approximation, this leads to the
following linear system for $c_1$ and $\rho$:
\begin{align}
  \label{eq:ssDiff}
  \Delta c_1 - \xi^{-1}\partial_x c_1 &= 0,  \\
  \label{eq:ssBc}
  c_1 - d_+\xi^{-1}c_0\cos\theta &=
  c^0_\mathrm{eq}\frac{\Gamma}{R_0^2}(\rho + \rho^{\prime\prime}), \\
  0 = \partial_t \rho &= -(v_1 - \vec{V} \cdot\vec{n}) \nonumber \\
 \label{eq:ss1}
 &=   a^2 D\big(\frac{c_1}{\xi} \hat{x}\cdot\vec{n}_0 -
  \nabla c_1 \cdot \vec{n}_0\big),
\end{align}
where $\vec{n}_0 = (-\cos\theta, \sin \theta )$ is the normal of the
circular steady state.

With the ansatz \eqref{eq:sep1} for $c_1$,
the steady state condition \eqref{eq:ss1} is equivalent to
\begin{align*}
  \label{eq:ss2}
  0 &= \partial_r f - \tfrac{1}{2\xi} f \cos \theta.
\end{align*}
Next, using  \eqref{eq:sep2} and property \eqref{eq:besselRec}
of the Bessel functions $I_n$ leads to the following
 simple recursion relation for the coefficients $\hat{c}_n$:
\begin{equation}
  \label{eq:rec}
  \hat{c}_{n-1} I_{n-1} +
  \hat{c}_{n+1} I_{n+1}
  = \hat{c}_{n} \big( I_{n-1} +
  I_{n+1} \big),
\end{equation}
where we have used the notation $I_m = I_m(R_0 / 2\xi)$.
For a solution that is symmetric under reflection at the 
$x$-axis (field direction) we have
$\hat{c}_n = \hat{c}_{-n}$, which together with \eqref{eq:rec} implies
that $\hat{c}_n \equiv \hat{c}_0$. Therefore any symmetric solution of
\eqref{eq:ssDiff},\eqref{eq:ss1} is of the form
\begin{align*}
  \label{eq:c1fourier}
  c_1(r,\theta) &=  \hat{c}_0 \exp(\tfrac{r}{2\xi}\cos\theta )  
 \sum_{n=-\infty}^{\infty}  I_n(\tfrac{r}{2\xi})\exp(in\theta)\\
  &= \hat{c}_0 \exp(\tfrac{r}{\xi}\cos\theta),
\end{align*}
where in the second identity we have used \eqref{eq:besselExp}.  Now the
boundary condition \eqref{eq:ssBc} is used to fix the constant $\hat{c}_0$ as
follows: First note that \eqref{eq:ssBc} describes a driven harmonic oscillator
in ''time'' $\theta$, with the left hand side being the driving force.
Next recall that a $2\pi$ periodic solution $\rho(\theta)$ of this
oscillator exists provided the driving force is  $2\pi$-periodic with
vanishing  $n=1$ Fourier mode, where the latter condition means that the oscillator is not
in resonance with the driving force.
Using the Fourier expansion of $c_1$ (see \eqref{eq:besselExp}) this determines
the constant $\hat{c}_0$ to be
\begin{equation*}
  \hat{c}_0 = d_+ \frac{c^0_\mathrm{eq}(1-\tfrac{\Gamma}{R_0})}{2\xi I_1(z)},
  \quad z = \frac{R_0}{\xi}.
\end{equation*}
Thus, the steady state adatom concentration is given by
\begin{equation*}
  c_1(r,\theta) =
  d_+ \frac{c^0_\mathrm{eq}(1-\tfrac{\Gamma}{R_0})}{2\xi I_1(z)} \exp(\tfrac{r}{\xi}\cos\theta).
\end{equation*}
Finally, the symmetric steady state shape ($\rho(\theta) = \rho(-\theta)$) is
obtained as the solution of the ordinary differential equation
\eqref{eq:ssBc} as:
\begin{align}
  \nonumber
  &\rho(\theta) = \rho_0(z) + \sum_{n\geq2} \rho_n(z) \cos n\theta \\
\label{eq:linSteadyState}
&= R_0  \delta z \frac{(\tfrac{R_0}{\Gamma} - 1)}{ I_1(z)}
  \Big(\tfrac{1}{2}I_0(z) + \sum_{n\geq 2}^\infty \frac{I_n(z)}{1 -
    n^2} \cos n\theta \Big).
\end{align}
The relative perturbation $\rho / R_0$ is expressed as a function of
the dimensionless parameters $\delta = d_+/R_0$, $z=R_0/\xi$ and $\Gamma/R_0$,
which characterize the deviation from the TD limit, the strength of the
electromigration force and the capillary effects, respectively.

\begin{figure}[htp]
  \includegraphics*[width=0.49\columnwidth]{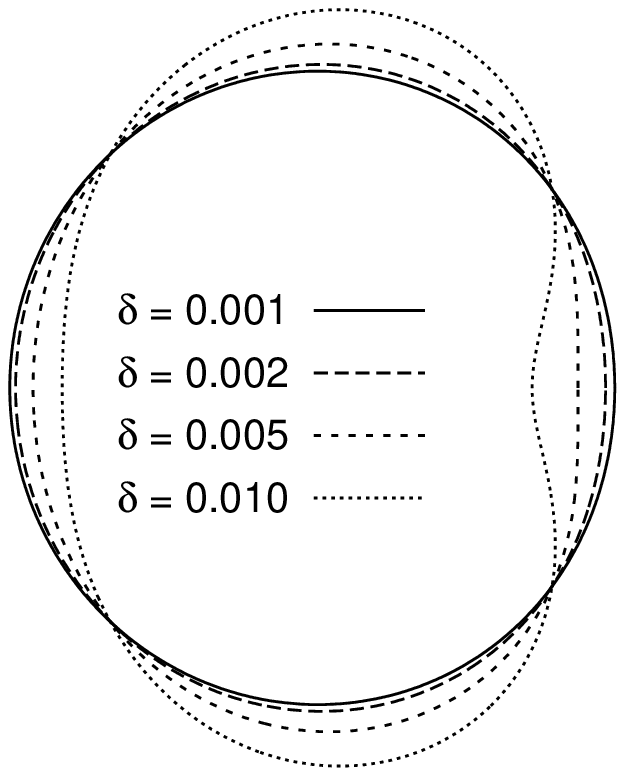}
  \includegraphics*[width=0.49\columnwidth]{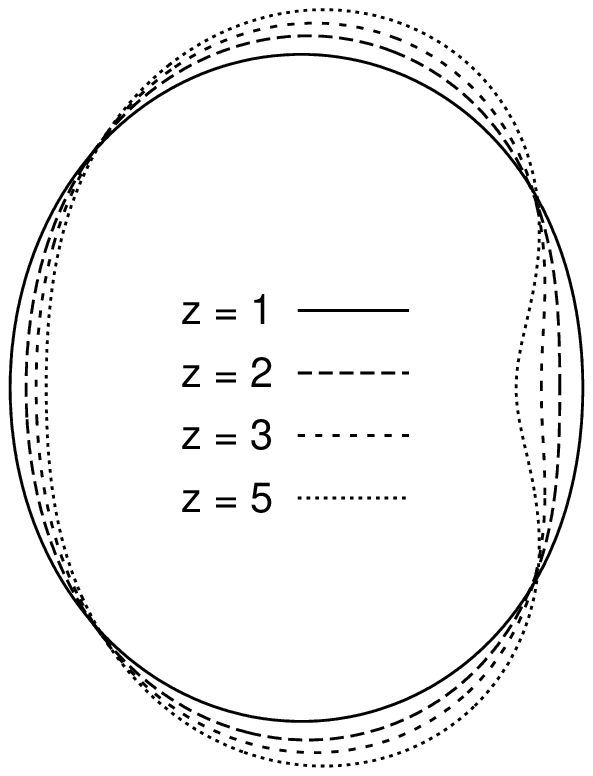}
  \caption{First order perturbations of the circular steady state.
    Left: Steady state shape for different values of $\delta = d_+/R_0 \ll 1$
    with fixed $z = R_0/\xi = 5$ and $\Gamma /R_0 = 0.05$. Right: Different
    values of $z$ with fixed $\delta z = 0.05$.  In both cases the perturbed
    shape $R_0 + \rho - \rho_0$ is depicted, i.e. the dilation mode has been subtracted.}
  \label{f:steadyStatesLinear}
\end{figure}

The constant term $\rho_0$ in (\ref{eq:linSteadyState}) describes a uniform
dilation of the circle. Since $\Gamma/R_0 \ll 1$, $\rho_0 > 0$, while
$\rho_{n \geq 2} < 0$. In particular, the leading order deformation with
$n = 2$ corresponds to an elongation perpendicular to the drift direction.
This is illustrated in Fig.~\ref{f:steadyStatesLinear}, where the dilation mode
has been subtracted. Moreover, with
increasing electromigration force (increasing $z$), the shapes start to become concave
on the trailing side (recall that the islands are moving to the left).  

As has been pointed out
above in Section \ref{s:model}, the full (nonlinear) evolution is area conserving
in the quasistatic limit. Since the
dilation mode $\rho_0$ is of the same order as the elongation mode $\rho_2$,
this property is generally violated by the first order perturbation
in $\delta$. This suggests that the perturbative regime
may be restricted to rather small elongations.
To obtain the steady states of the full nonlinear model, numerical simulations
of the time-dependent equations \eqref{eq:diff}-\eqref{eq:vel} have been performed.  
An adaptive finite element method is used, where the free boundary problem is discretized
semi-implicitly using an operator splitting approach and two independent
numerical grids for the adatom density $c$ and for the island boundary,
respectively.  For the boundary evolution, a front tracking method is applied,
for details see \cite{BHLLV2004}.

\begin{figure}[htbp]
  \includegraphics*[width=1.\columnwidth]{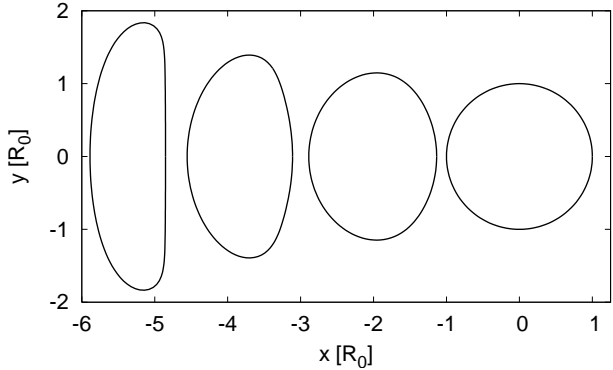}
  \caption{Simulation of the full nonlinear evolution of a circular vacancy island
    towards a steady state.  For this example, $\delta = 0.1$ and $z = 5$,
    which is well beyond the regime of validity of the first order perturbation
    theory.  The void is moving from right to
    left and the snapshots are taken at $t = 0$, $t=1.5$, $t=5$,
    $t=50$.  The contours at later times are moved back to the right for
    better visibility. Space is measured in units of the radius $R_0$ of the
    initial shape and time in units $t_0 = R_0/|\vec{V}|$, where
    $\vec{V}$ is the drift velocity of a circular island
    as given in \eqref{eq:velDrift}. The simulation parameters are
    $R_0 = 100a, \; a^2 c^0_\mathrm{eq} = 10^{-5},\; \Gamma/R_0 = 0.05$.
}
  \label{f:steadyStatesFEM1}
\end{figure}

\begin{figure}[htbp]
  \includegraphics*[width=0.7\columnwidth]{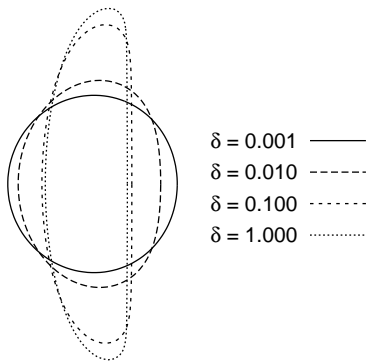}
  \caption{Steady state shape as approached by the nonlinear evolution
    for different values of $\delta = d_+/R_0 \ll 1$
    with fixed $z = R_0/\xi = 5$ and $\Gamma /R_0 = 0.05$.  }
  \label{f:steadyStatesFEM2}
\end{figure}

Starting with a circular shape, the void
elongates until it reaches a steady state.  A typical example of the time
evolution is depicted in Fig.~\ref{f:steadyStatesFEM1}.
Here the parameters are $\delta = 0.1$, $z = 5$, which, as will be seen below,
is already far away from the perturbative regime.  We do not observe any concave
parts at the back side of the void as opposed to the perturbative steady state, see
Fig.~\ref{f:steadyStatesLinear}.  This turns out to be true for all examples
which we investigated.  We also checked that the final steady state does not
depend on the initial shape. For example, starting with a "bean-like" shape being the
steady state shape as obtained from the perturbation theory also leads to the
same convex steady state.  Moreover, we have not seen any break up in the
simulations even for cases where $\delta \sim 1$.  In all cases, a steady
state shape similar to the one depicted in Fig.~\ref{f:steadyStatesFEM1} is
approached, where the deformation increases with increasing $\delta$ and where
the curvature at the back side (right side) approaches zero.  In
Fig.~\ref{f:steadyStatesFEM2} the steady states as obtained from the nonlinear
evolution are depicted for different values of $\delta$.
By comparing with Fig.~\ref{f:steadyStatesLinear} (left) it is clear that the
perturbative regime is limited to rather small values of $\delta$, since already for
$\delta = 0.01$ the perturbative (non-convex) and the nonlinear (convex) steady state
shapes differ considerably.

\begin{figure}[hbtp]
  \includegraphics*[width=0.9\columnwidth]{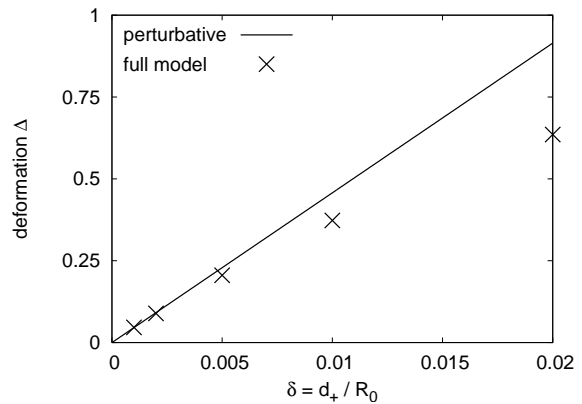}
  \caption{Deformation $\Delta$ of the steady state shape versus
    relative kinetic length $\delta = d_+/R_0$, as obtained from perturbation
    theory and from the full model.}
  \label{f:steadyStatesFEM3}
\end{figure}

In Fig.~\ref{f:steadyStatesFEM3} we have investigated this quantitatively by
comparing the deformation $\Delta$
\begin{equation*}
  \Delta = \big( \rho(\pi / 2) + \rho(-\pi/2) - \rho(0) - \rho(\pi) \big) / 2R_0
\end{equation*}
for the steady state obtained by the perturbation theory, i.e. given in
\eqref{eq:linSteadyState}, and for the steady state as approached by
simulations of the full nonlinear dynamics. The first order perturbation 
theory is seen to be quantitatively accurate only for $\delta < 0.01$. 
%
%
%
%
%
%
%
\section{Linear stability analysis}
\label{s:linStability}
Next we perform a linear stability analysis for the circular steady state
of the interior model in the TD limit.
Thus we are looking for small perturbations of the form
\begin{align*}
  \rho(\theta,t) &= \rho(\theta,0) \exp( i\omega t), \\
  c_1(r,\theta, t) &= c_1(r,\theta,0) \exp( i\omega t).
\end{align*}
As in \eqref{eq:ssDiff}-\eqref{eq:ss1}
we obtain -- using the quasi-static approximation for $c_1$ --
 the following linearized system for $c_1$ and $\rho$,
\begin{align}
  \label{eq:linDiff}
  \Delta c_1 &- \xi^{-1}\partial_x c_1 = 0,  \\
  \label{eq:linBc}
  c_1  &=  c^0_\mathrm{eq}\frac{\Gamma}{R_0^2}(\rho + \rho^{\prime\prime}),\\
\label{eq:linVel}
 \partial_t \rho &=
   a^2 D\big(\frac{c_1}{\xi} \hat{x}\cdot\vec{n}_0 - \nabla c_1 \cdot \vec{n}_0\big).
\end{align}
In view of \eqref{eq:sep1} we make the following ansatz
\begin{align}
  \label{eq:ansatzRho}
  \rho(\theta,0) &= \exp(\tfrac{z}{2}\cos\theta) \sum_n
  \rho_n\exp(in\theta) , \quad z = \frac{R_0}{\xi}.
\end{align}
Thus, the perturbation $\rho(\theta,0)$ is written as a series expansion in
the functions $\exp(\tfrac{z}{2}\cos\theta)\exp(in\theta)$, instead of the
usual Fourier modes.  This choice has the advantage of simplifying the
subsequent calculation. In particular it will lead to a linear system, where
each matrix row/column has only a small number of non zero entries. However,
as we will discuss later, the non-orthogonality of the ansatz functions
increases the non-normality of the matrix.  Equation \eqref{eq:linVel} now
takes the form
\begin{equation*}
  i\omega \sum \rho_n \exp(in\theta) =
  a^2 D \big(\partial_r f - \tfrac{1}{2\xi} f \cos \theta \big).
\end{equation*}
Using \eqref{eq:sep2}, \eqref{eq:besselRec} and matching coefficients
we obtain
\begin{align}
\nonumber
  i \omega \rho_n = & \frac{a^2 D}{4\xi} \Big(\hat{c}_{n} \big(
  I_{n-1} + I_{n+1} \big)\\
  \label{eq:linVel2}
  &\qquad -
  \hat{c}_{n-1} I_{n-1} -  \hat{c}_{n+1}  I_{n+1} \Big),
\end{align}
where here and in the following we use the notation
$I_n = I_n(\tfrac{z}{2})$.
We will finally use the linearized boundary condition \eqref{eq:linBc}
to express the coefficients $\hat{c}_n$ in \eqref{eq:linVel2} in
terms of the $\rho_n$'s.  Inserting the ansatz
\eqref{eq:sep1} for $c_1$, and \eqref{eq:ansatzRho} for $\rho$ into
\eqref{eq:linBc} yields
\begin{align*}
  \nonumber
  f(R_0,\theta) &= c^0_\mathrm{eq}\frac{\Gamma}{R_0^2}\sum\Big( 1 +
  \frac{z^2}{8}(1 - \cos 2\theta)
  - \frac{z}{2}\cos \theta \\
  &\qquad- inz\sin\theta - n^2 \Big) \rho_n \exp(in\theta),
\end{align*}
which leads to  (using \eqref{eq:sep2} for the left hand side)
\begin{align}
  \nonumber
  \hat{c}_n I_n(\tfrac{z}{2}) &= c^0_\mathrm{eq}\frac{\Gamma}{R_0^2}\Big(
  \big(1 + \frac{z^2}{8} - n^2\big)\rho_n -
  \frac{z^2}{16}\big(\rho_{n+2} + \rho_{n-2}\big) \\
  \label{eq:cn}
  &\qquad- \frac{z}{4}\big( (2n-1)\rho_{n-1} - (2n+1)\rho_{n+1} \big) \Big).
\end{align}
Inserting \eqref{eq:cn} into \eqref{eq:linVel2} leads after some
tedious but straight forward calculations to the following eigenvalue
problem for the coefficients $\{\rho_n\}$ and $\omega$:
  \begin{align}
  \label{eq:evProblem}
   \nonumber \lambda \rho_n &= \Big( \frac{z}{2}\big(1 +
  \frac{z^2}{8} - n^2\big)C_n + \frac{z^2}{8} \Big) \rho_n \\
  \nonumber &+ \Big( \frac{z}{4}n(n+2) + (2n+1)\frac{z^2}{8}C_n -
  \frac{z^3}{64} \Big) \rho_{n+1} \\ \nonumber
  &+ \Big( \frac{z}{4}n(n-2) + (-2n +
  1)\frac{z^2}{8}C_n - \frac{z^3}{64} \Big) \rho_{n-1} \\  &-
  \Big( \frac{z^2}{16}(2n+3) +\frac{z^3}{32}C_n \Big) \rho_{n+2} \\
  \nonumber &+
  \Big( \frac{z^2}{16}(2n-3) -\frac{z^3}{32}C_n \Big) \rho_{n-2} \\
  \nonumber &+ \frac{z^3}{64}\big( \rho_{n+3} + \rho_{n-3} \big),
\end{align}
with 
$$
\lambda = \frac{ i\omega R_0^3}{a^2c^0_\mathrm{eq}D\Gamma}, \;\;\; 
C_n =
\frac{I_{n+1} + I_{n-1}}{ 2 I_n} =
\frac{I_{n+1}}{I_n} + \frac{2n}{z}.
$$
The right hand side of (\ref{eq:evProblem}) 
represents the linearized time evolution of the
island as a (infinite) matrix $\hat{A}$ acting on the coefficients
$\rho_n$. In real space it corresponds to an integro-differential operator,
which is essentially nonlocal. Remarkably, the matrix depends on the system
parameters only through the dimensionless electromigration force $z = R_0/\xi$. 
In particular, the capillary length 
$\Gamma$ affects the time scale of the linear evolution, but not
the stability of specific perturbations \cite{PierreLouisEinstein2000}.
For $\Gamma \to 0$, all eigenfrequencies $\omega_n$ vanish, which implies that
all perturbations become marginal. Indeed, it is easy to check that
in the TD limit, for $\Gamma = 0$, \textit{any} island shape translates rigidly   
at constant velocity under \eqref{eq:diff}--\eqref{eq:vel}. 

The matrix $\hat{A}$ exhibits a symmetry, which
originates from the invariance of the system under reflection at the $x$-axis
(field direction). In terms of the coefficients $\rho_n$ this reflection is
expressed as $\rho_n \to \rho_{-n}$, and one readily verifies that this does
leave (\ref{eq:evProblem}) unchanged. Accordingly the eigenspace splits
into two invariant subspaces with symmetric and antisymmetric eigenmodes
characterized by
\begin{align*}
  \rho_{n} &= \rho_{-n}    &\mathrm{symmetric}  \\
  \rho_{n} &= -\rho_{-n}  &\mathrm{antisymmetric}.
\end{align*}
In both cases the eigenmodes are fully determined by only half of the
coefficients (e.g. those with positive index), which allows us to reduce
(\ref{eq:evProblem}) to a semi-infinite system. By truncating it towards large
$n$ (cutoff towards small wavelengths) we arrive at a finite linear system
which we solve numerically.

Before we present the numerical results, we discuss {\em translations} and
{\em dilations}, which are perturbations related to the symmetry properties
of the system. Symmetry under translations in the horizontal ($x$) and vertical
($y$) direction leads to two zero eigenmodes $\mathcal{T}_x$, $\mathcal{T}_y$
-- the infinitesimal horizontal and vertical translations -- given by
\begin{align*}
  \mathcal{T}_x(\theta)=\cos(\theta), \quad
  \mathcal{T}_y(\theta)=\sin(\theta).
\end{align*}
Indeed, inserting $ \rho = \mathcal{T}_x$ or $\rho = \mathcal{T}_y$ into the
linearized boundary condition (\ref{eq:linBc}) leads to $c_1\equiv0$ and
therefore (\ref{eq:linVel}) yields $\partial_t \rho \equiv 0$. 
The horizontal translation $\mathcal{T}_x$ belongs to the symmetric eigenmodes,
while the vertical translation $\mathcal{T}_y$ belongs to the antisymmetric class. 
Next we consider a {\em dilation} $\mathcal{D}$, i.e., a constant initial
perturbation
\begin{equation*}
  \mathcal{D}(\theta) = 1.
\end{equation*}
Inserting $\rho(\theta)= \mathcal{D}(\theta) = 1$ into the boundary condition
(\ref{eq:linBc}) yields
\begin{equation*}
  c_1 =\frac{c_\mathrm{eq}^0 \Gamma}{R_0^2},
\end{equation*}
which reflects the fact that one passes from one stationary solution to
another by increasing the radius of the island and the concentration inside
the island by a constant value.  However, a dilation is {\em not} a zero
eigenmode: Since we consider perturbations of a circular steady state, which
has a steady state drift velocity depending on the radius, two circles with
different radius are drifting apart. This leads to a linear increase of the
perturbation. From \eqref{eq:linVel}, an initial perturbation
$\rho(\theta,0) = 1$ has to grow according to
\begin{equation*}
   \partial_t \rho = \frac{a^2 c_\mathrm{eq}^0 \Gamma D}{R_0^2 \xi} \cos(\theta)
   =  \frac{a^2 c_\mathrm{eq}^0 \Gamma D}{R_0^2 \xi} \mathcal{T}_x(\theta).
\end{equation*}
In that sense a dilation $\mathcal{D}$ generates a translation, and
$\mathcal{D}$ is a generalized eigenmode with eigenvalue zero according to
$\hat{A}^2 \mathcal{D} \sim \hat{A} \mathcal{T}_x = 0$. Thus (restricting to
the symmetric case) the eigenvalue zero is two-fold degenerate, and has one
(proper) eigenvector. Therefore the matrix $\hat{A}$ can not be diagonalized
completely but contains a $2\times2$ Jordan block corresponding to the
invariant subspace spanned by $\mathcal{D}$ and $\mathcal{T}_x$. Apart from
that, the dilation doesn't play a role, because the time evolution preserves
the area and we can therefore always restrict ourselves to perturbations which
do not contain dilations.

\begin{figure}[htp]
  \includegraphics*[width=0.5\textwidth,angle=0]{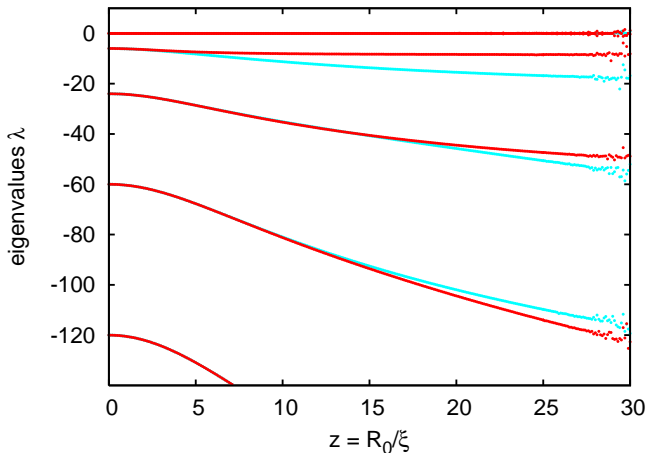}
  \caption{(Color online) Spectrum of the linearized theory as obtained by numerical
    solution of the eigenvalue problem \eqref{eq:evProblem}.  Depicted are the
    eleven largest eigenvalues as a function of the dimensionless electromigration force $z =
    R_0 / \xi$. Eigenvalues of symmetric modes are red/black, eigenvalues of
    antisymmetric modes blue/grey. Note that the eigenvalue zero is threefold
    degenerate. All negative eigenvalues come in pairs, consisting of a symmetric
    and an antisymmetric mode which become degenerate at $z=0$.
   For the fluctuations appearing around $z=30$ see the
    discussion at the end of Section \ref{s:linStability}.
}
  \label{f:spectrum}
\end{figure}
The numerically determined spectrum is presented in Fig.~\ref{f:spectrum}.
Here the largest eigenvalues, for $0\leq z \leq 30$  are depicted.
The spectrum is purely real and apart from the predicted
threefold degenerate zero eigenvalue it is strictly negative.
For $z=0$ (no electromigration force), the right hand side of
\eqref{eq:evProblem} becomes diagonal:
\begin{equation*}
  \lambda \rho_n = |n|(n^2-1)\rho_n,
\end{equation*}
which allows to directly read off the eigenvalues $\lambda_n = |n|(n^2-1)$.
Each eigenvalue $\lambda_n$ is twofold degenerate with eigenmodes given by the
Fourier modes $\cos(n\theta)$, $\sin(n\theta)$.  Since no field breaks the
rotational symmetry, the symmetric and antisymmetric modes $\cos(n\theta)$,
$\sin(n\theta)$ are connected by a rotation by $\pi/2$ and belong to the same
eigenvalue $\lambda_n$.  In the presence of an electric current, i.e. $z>0$, the
rotational symmetry is broken and the degeneracy is removed, i.e., the
eigenvalues split. Moreover increasing values of $z$ lead to decreasing
eigenvalues, i.e. larger islands or islands in the presence of a stronger
field relax faster to the circular shape (as compared to the case
without drift). For large values of $z$ the eigenvalue
problem becomes numerically difficult to solve, leading to a noisy spectrum in
Fig.~\ref{f:spectrum}
 for values of $z \sim 30$. This will be discussed in more detail below.

\begin{figure}[htp]
  \includegraphics*[width=0.5\textwidth,angle=0.0]{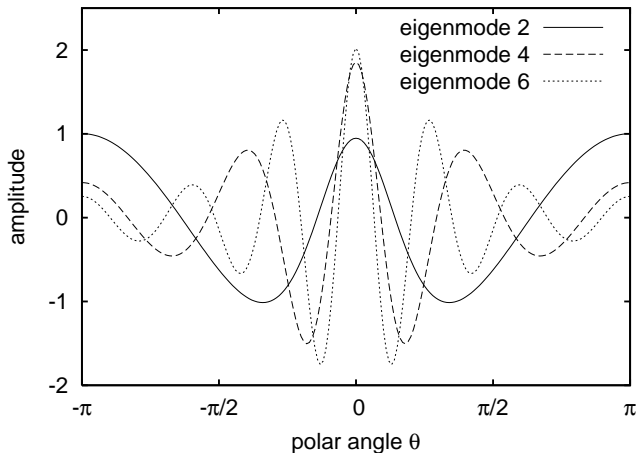}
  \caption{Second, fourth and sixth symmetric eigenmode as function
    of the angle $\theta$ ($z=10$). $\theta=\pi$ is the drift
    direction. The modes are normalized with respect to the $L^2$-norm.}
  \label{f:eigenmodes1}
\end{figure}

\begin{figure}[htp]
\includegraphics*[width=0.32\columnwidth,angle=0]{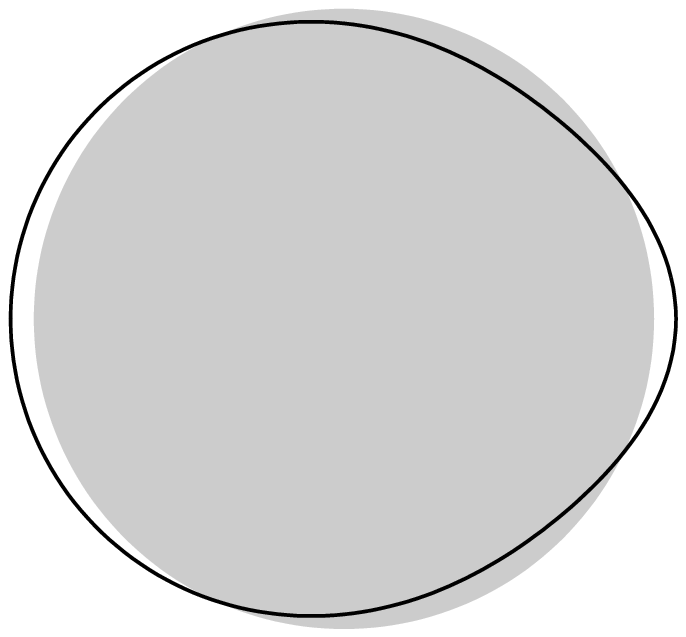}
\includegraphics*[width=0.32\columnwidth,angle=0]{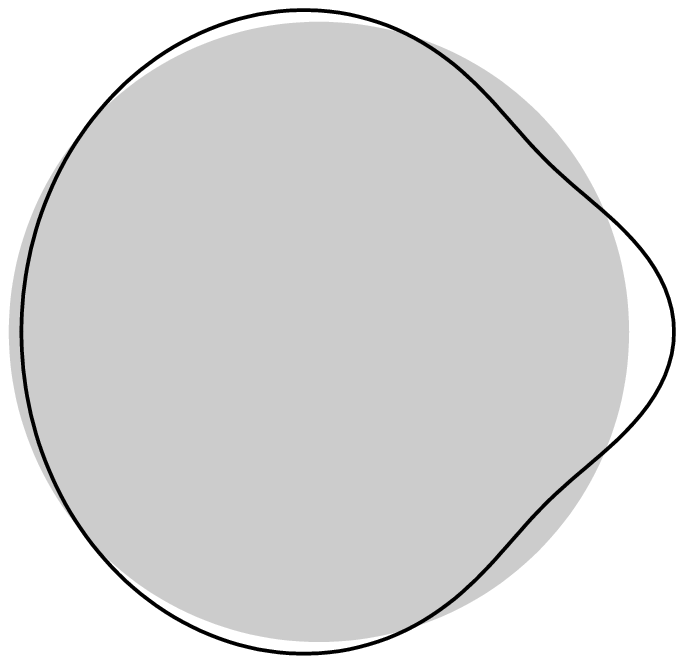}
\includegraphics*[width=0.32\columnwidth,angle=0]{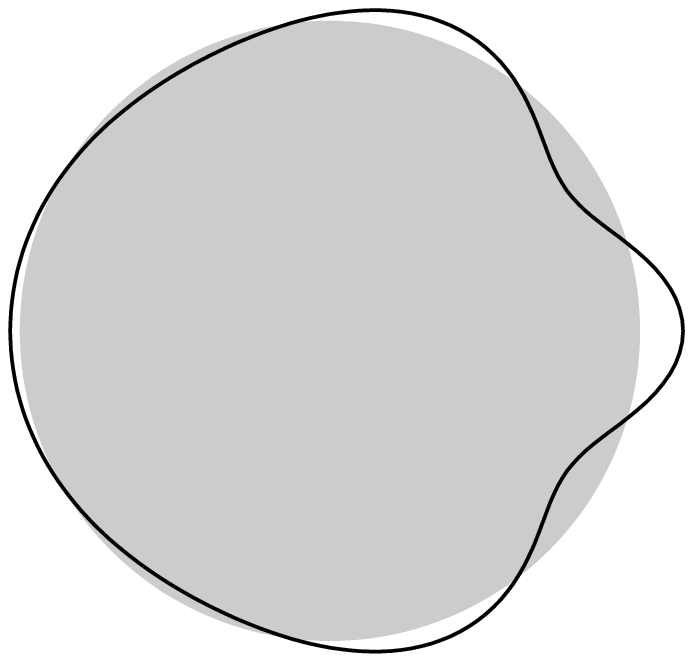}\\
\vspace{0.2cm}
\includegraphics*[width=0.32\columnwidth,angle=0]{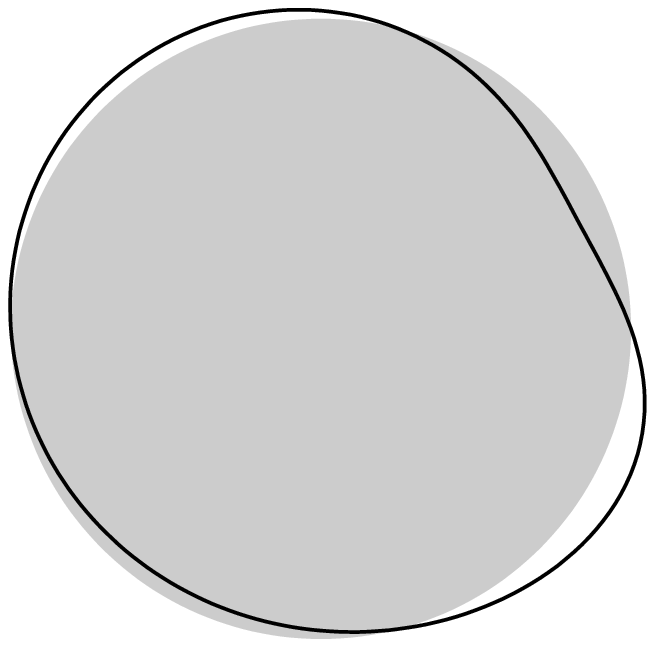}
\includegraphics*[width=0.32\columnwidth,angle=0]{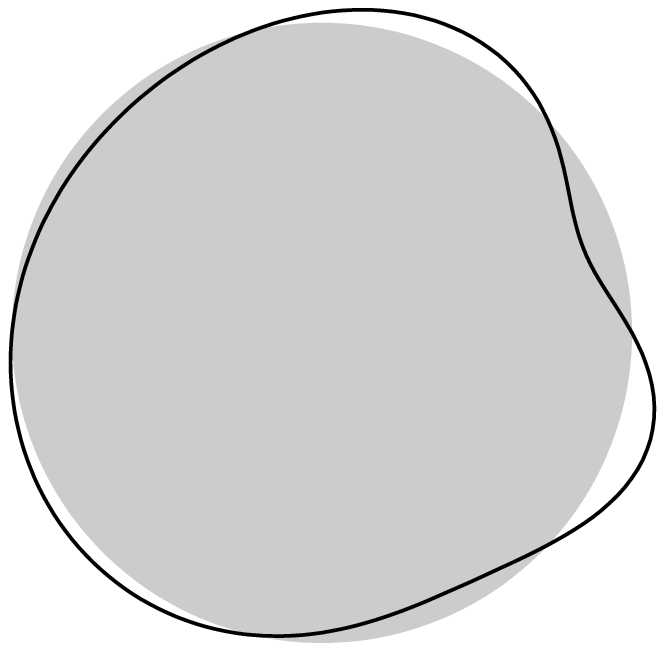}
\includegraphics*[width=0.32\columnwidth,angle=0]{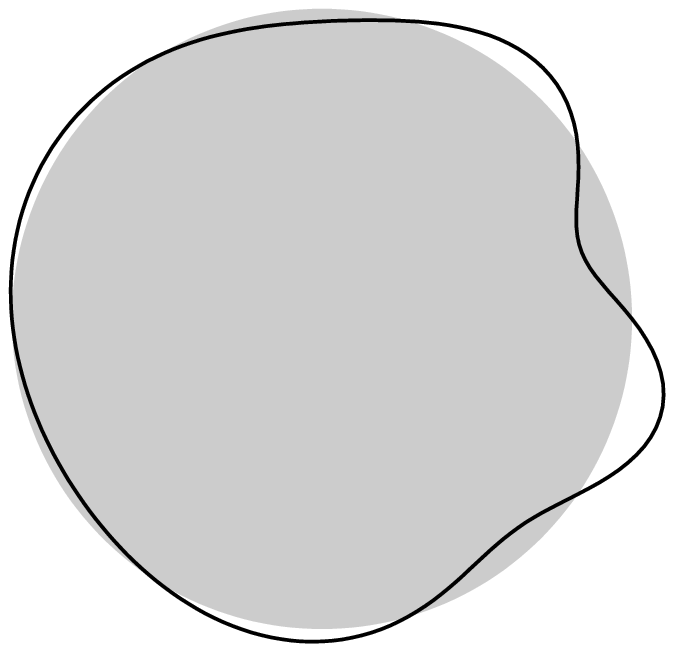}
\caption{Second, third and fourth eigenmode superimposed on a circle (grey background)
   for $z=10$. Top (bottom) row shows symmetric (antisymmetric) modes.}
  \label{f:eigenmodes2}
\end{figure}

\begin{figure}[htp]
  \includegraphics*[width=0.5\textwidth,angle=0.0]{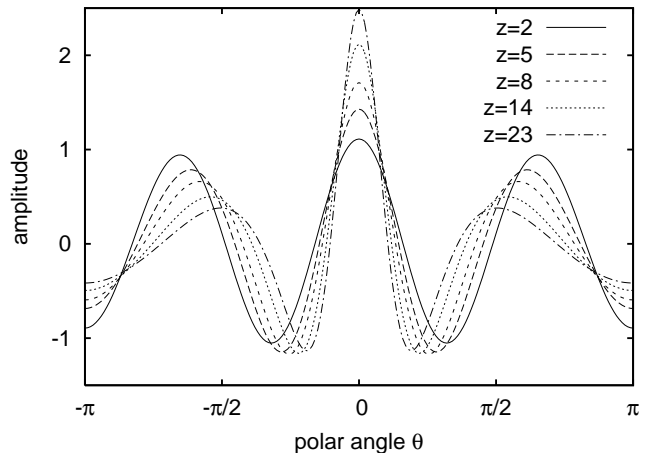}
    \caption{Symmetric eigenmode $n=3$  for different values of $z$. The asymmetry with
    respect to front ($\theta=\pi$) and back side ($\theta = 0$) increases
    with increasing value of $z$. The modes are normalized with respect to the $L^2$-norm.}
  \label{f:eigenmodes5}
\end{figure}

We now turn to the eigenmodes of the linearized time evolution.
Figs.~\ref{f:eigenmodes1} and \ref{f:eigenmodes2}
show some examples of symmetric and antisymmetric eigenmodes with small index
$n$ for $z=10$. All of them reveal a typical undulated shape, where the
number of nodes increases with the index and the amplitude is largest towards
$\theta=0$. Thus the modulation is more pronounced at the back side of the island
(with respect to the drift motion) and, as shown in Fig.~\ref{f:eigenmodes5}, 
this localization gets stronger with increasing $z$.
This is a signature of the fact that the eigenvectors become
more and more parallel (see below). The
increasing localization of the eigenmodes at the back side of the island may
be traced back to the convective, electromigration-induced flux in the adatom
diffusion equation~\eqref{eq:diff} which leads to the factor
$\exp(\tfrac{z}{2}\cos\theta)$ in \eqref{eq:ansatzRho}.  For large values
of $z$, this factor strongly suppresses all contributions for values of
$\theta$, which are not close to $\theta=0$. In particular, a perturbation
that is front-back symmetric has to be a linear combination of many different
eigenmodes.  In that case, the eigenmodes with large index $n$ will decay very
fast leading to a shape being close to the steady state circular shape at the
front side, while still having a large buckle at the back side.

This behaviour will be investigated in more detail in the next section, when
we consider the fully nonlinear evolution.  For large values of $z$ this will
finally lead to a fingering instability at the back side of the island.

Before turning to the nonlinear evolution, we comment on some observations
connected with non-normal eigenvalue problems.  The matrix $\hat{A}$ in
\eqref{eq:evProblem} is highly non-normal for large values of $z$, which
leads to a strong sensitivity of the spectrum with respect to small
perturbations of the matrix entries. Small changes of the entries of the
matrix may lead to structurally completely different spectra, a feature which
can be consistently described within the theory of pseudospectra
\cite{Trefethen2005}. The degree of non-normality depends on the choice of
the basis. In our case the basis is not orthogonal (with respect to the usual
$L^2$ scalar product) and therefore, although convenient for the analytical
part, can give a distorted picture of the underlying geometry.  We therefore
checked the results by transforming the system \eqref{eq:evProblem} to the
Fourier basis. Here the non-normality is much less pronounced and we obtained
the same spectrum as depicted in Fig.~\ref{f:spectrum}. In both cases, we have
not been able to obtain the spectrum numerically for values $z > 30$ since, as
can be seen from Fig.~\ref{f:spectrum} the lines start to become noisy between
$z=25$ and $30$. The fluctuations are getting stronger very quickly, making it
impossible to determine the spectrum reliably.

A closer examination reveals that the
condition numbers for the eigenvalues grow exponentially with $z$ and
therefore the accuracy of the results gets lost very
quickly at some point. This is again a consequence of the non-normality of the
problem. The condition number of a given eigenvalue is large when the
corresponding left and right eigenvectors are almost
orthogonal \cite{Quarteroni2000}. This is intimately connected with the
non-orthogonality between different (right) eigenvectors. In our case the
eigenvectors become almost parallel for large values of $z$. Even a slight perturbation
of the matrix entries will destroy this and improve the
condition numbers dramatically, of course by changing the spectrum
beyond recognition.

Finally we note that non-normality has also been discussed
\cite{Schimschak1998,Trefethen1993,Grossmann2000,Trefethen2005} as a mechanism causing a
transient growth of small perturbations even in the case of a linearly stable
system. This transient amplification can be large enough to drive the system
out of the linear regime, this way leading to an instability.  We thoroughly
investigated the transient behaviour of the linearized system and found that
this scenario does \textit{not} apply to our case.
%
%
%
%
\section{Nonlinear evolution}
\label{s:nonLinEv}
To further investigate the nonlinear evolution, we have performed numerical
simulations of \eqref{eq:diff}-\eqref{eq:vel} using
 an adaptive finite element method, as described in Section~\ref{s:steadyStates}.
We note that adaptive mesh refinement during the time evolution in regions with high curvature
turned out to be crucial for accurate simulations in the case when small
fingers appear in the shape.

\begin{figure}[tp]
  \includegraphics*[height=0.45\textheight]{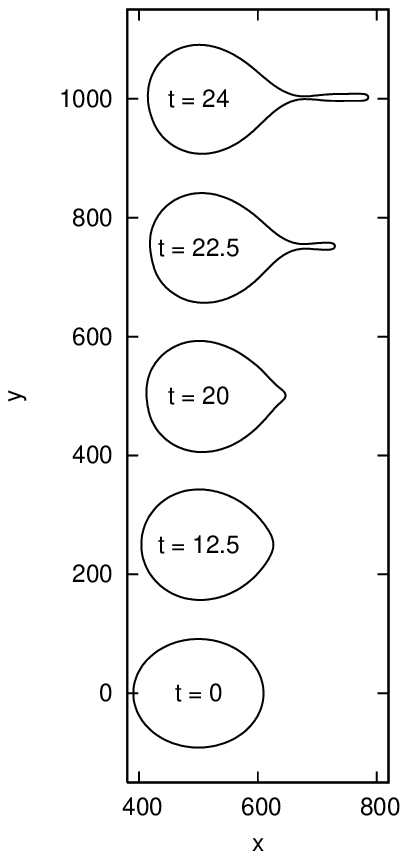}
  \includegraphics*[height=0.45\textheight]{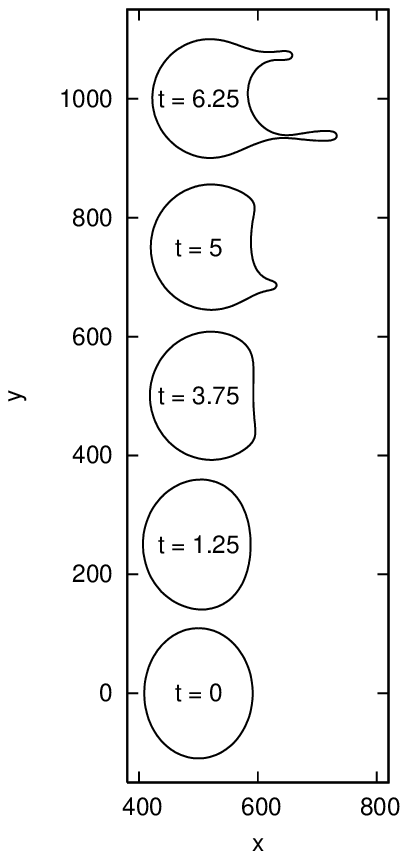}
  \caption{Nonlinear dynamics in the interior model with strong
    electromigration force ($z=25$). The initial shape is an ellipse with
    aspect ratio $1.2$. A fingering instability appears at the trailing end,
    which leads to a pinch-off.
    The contours are shifted upwards for better visibility and also shifted to the right with
    $\Delta x = t |\vec{V}|$, where $\vec{V}$ is the drift velocity of a
    circular island, see \eqref{eq:velDrift}.
    Space is measured in units of the lattice spacing $a$ and time in
    units $t_0 = R_0/|\vec{V}|$. The simulation parameters are $R_0 = 100a, \;
    a^2 c^0_\mathrm{eq} = 10^{-5},\; \Gamma/R_0 = 0.02$.   
    The breaking of the up-down symmetry in the right column is due to
    numerical noise.}
  \label{f:nonlin1}
\end{figure}

We probe the nonlinear dynamics with a
very  strong electromigration force, i.e. choosing $z = R_0/\xi = 25$.
In Fig.~\ref{f:nonlin1} a simulation of a vacancy island (interior
model) in the TD regime is depicted.  Here an ellipse with aspect ratio $1.2$
has been taken as the initial shape.
We observe a fingering instability at the back side of the
vacancy island, which finally leads to a pinch-off. Note that this behaviour
has not been predicted by the linear stability theory.

\begin{figure}[tp]
  \includegraphics*[width=0.45\textwidth]{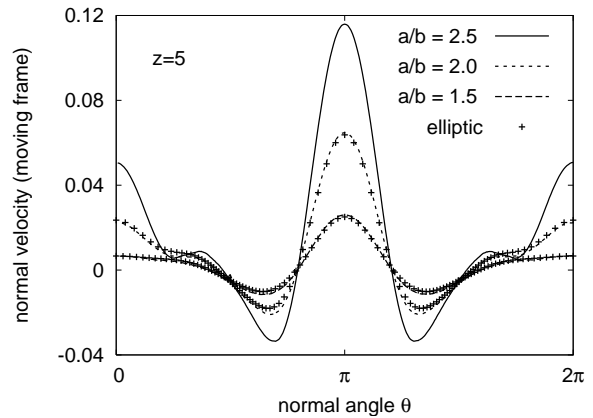}
  \caption{Normal velocity of an ellipsoidal vacancy island (elongated in
    the horizontal direction) as a function of the normal angle
    in the center of mass system, i.e., after subtracting the drift velocity
    of the island.  The normal points inwards, i.e. a positive velocity
    corresponds to local shrinking, and the angle $\theta = \pi$ corresponds
    to the drift direction.  The normal velocity taken from the numerical
    simulations by averaging over the first $1000$ time steps is in very good
    agreement with the results obtained by evaluating the analytic solution
    using elliptic coordinates.}
  \label{f:normalVel}
\end{figure}

To gain a better understanding of the nonlinear dynamics, the initial
evolution of an ellipsoidal island has been investigated in more detail. For
this special geometry, one may use elliptic coordinates and an expansion of
the adatom concentration in terms of Mathieu functions to arrive at an
analytic expression for the normal velocity, see Appendix~\ref{a:Mathieu},
\eqref{eq:velEllipt}.  However, as it turned out, a reliable evaluation of
the series expansion is possible only for moderate values of $z \leq 5$ and
aspect ratios $\leq 2$.  For parameters in this regime, the normal velocity of
an ellipsoidal vacancy island -- elongated in the horizontal direction -- in
the center of mass system, i.e. after subtracting the (fairly large) drift
velocity, is compared with the corresponding results of the finite element
simulation in Fig.~\ref{f:normalVel}.

\begin{figure}[tp]
  \includegraphics*[width=0.49\textwidth]{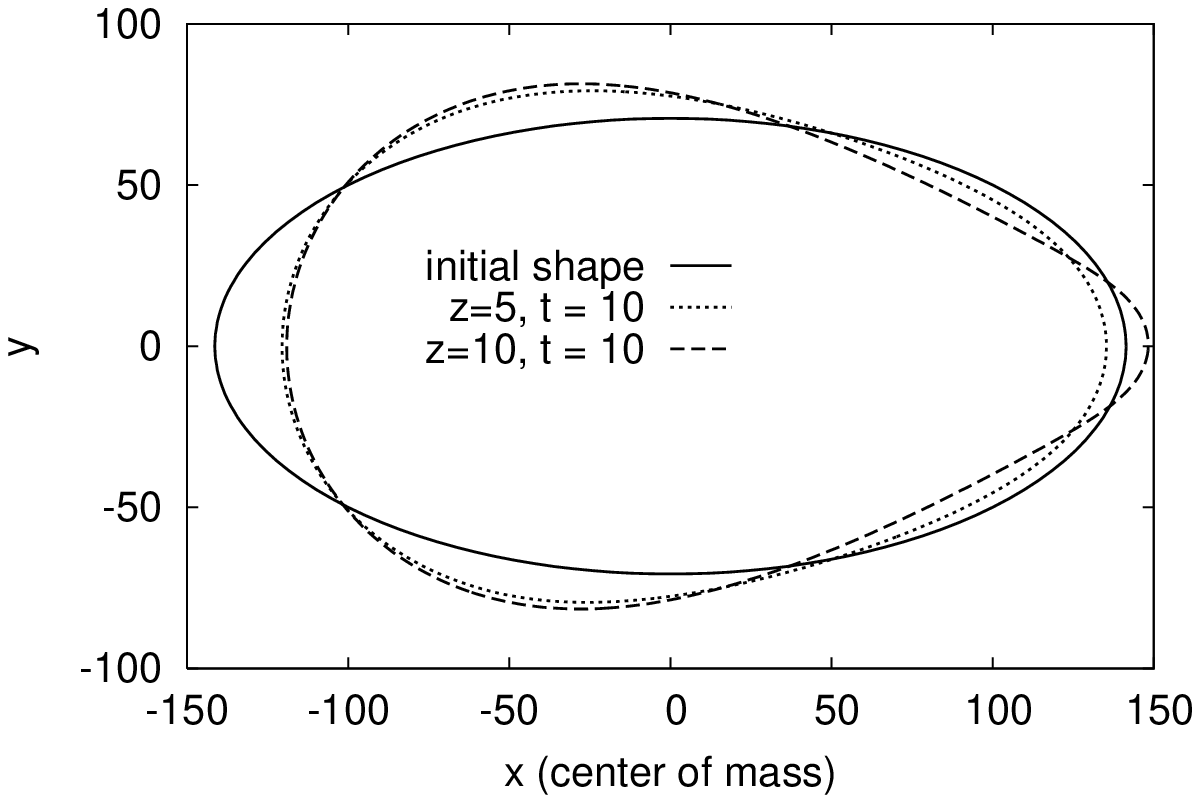}
  \includegraphics*[width=0.49\textwidth]{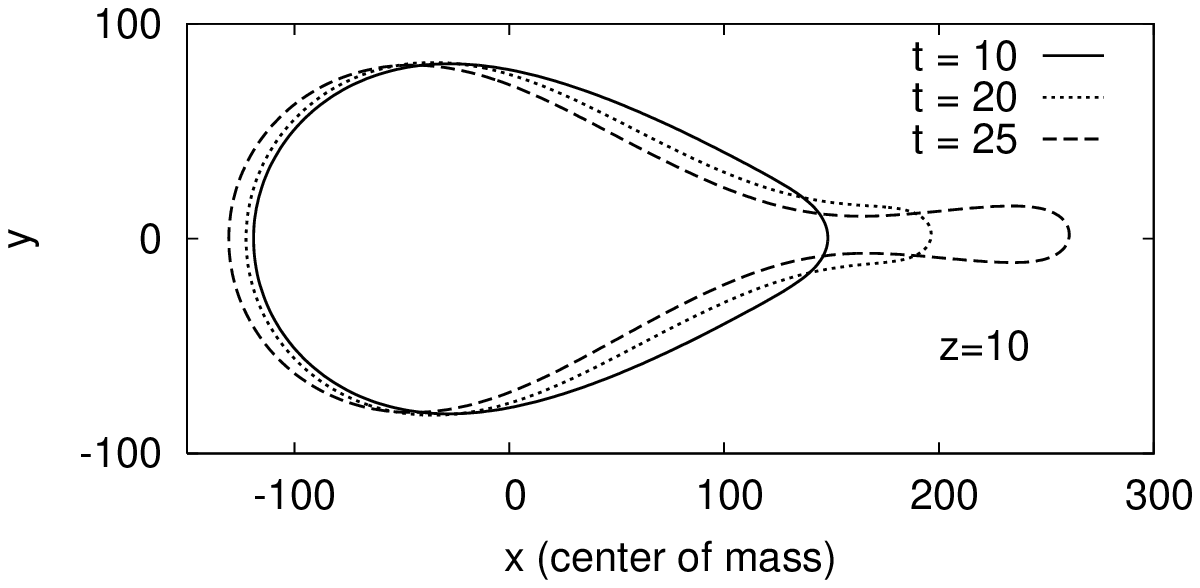}
  \caption{Numerical simulation of the evolution of an ellipsoidal vacancy
    island with aspect ratio $2.0$ and area $\pi R_0^2$.
    The upper figure depicts the initial stage of the evolution:
    the left/right asymmetry increases with increasing value of $z=R_0/\xi$,
    leading to the formation of an egg-like shape.
    Bottom: For large values of $z$, the asymmetric evolution leads to a
    fingering instability at the back side of the island.
    Space is measured in units of the lattice spacing $a$ 
    and time in units $t_0 = R_0/|\vec{V}|$, where
    $\vec{V}$ is the drift velocity of a circular island 
    as given in \eqref{eq:velDrift} for the case $z=10$. 
    The simulation parameters are
    $R_0 = 100a, \; a^2 c^0_\mathrm{eq} = 10^{-5},\; \Gamma/R_0 = 0.05$.
    }
  \label{f:eggFormation}
\end{figure}

The two solutions are in very good agreement. Moreover, one realizes that the
dynamics is fastest at the front part of the  island and tends to relax the front
half  of the ellipse towards a circle, since the boundary moves inwards at
$\theta = \pi$ ( front ) but outwards in nearby regions with $\theta \approx
\pi \pm \pi/4$. Since the evolution at the back half of the island is slower
and always inwards, we expect the dynamics to lead to an egg like shape, which
is verified in the simulations, see Fig.~\ref{f:eggFormation}. Moreover, the
asymmetry of the dynamics becomes more pronounced with increasing values of
$z$. We expect this to be a possible reason for the onset of a fingering
instability, once a critical value of the curvature at the back end is reached.

\begin{figure}[tp]
  \includegraphics*[width=0.45\textwidth]{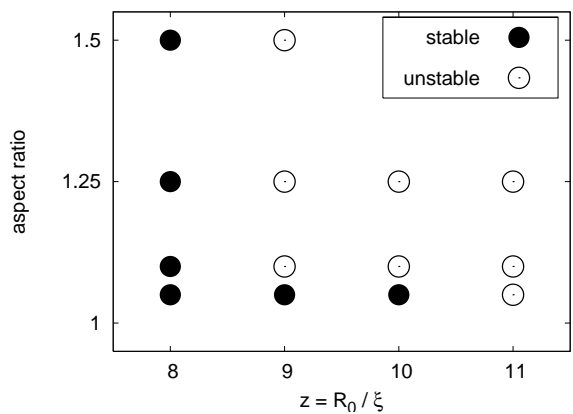}
  \caption{Threshold for the onset of the instability:
    Simulations for different values of $z=R_0/\xi$ and aspect ratios of
    the initial ellipsoid, where the elongation is in field direction, have
    been performed. The instability sets in at $z = 9$.  For $z\geq 11$
    also the smallest aspect ratio leads to an instability, i.e. a fingering
    at the back side of the island.}
  \label{f:stabilityDiagram}
\end{figure}

To roughly locate the
threshold for the onset of the instability we systematically varied
the electromigration strength $z$ and the amplitude of the 
initial perturbation given as
the aspect ratio of the initially horizontally elongated circle.  The results
of the simulations are presented in Fig.~\ref{f:stabilityDiagram}.  As can be
seen, the instability sets in around $z \sim 10$.

Although we are so far lacking a satisfactory analytic understanding
of the underlying mechanism, 
we would like to present some intuitive reasoning
for the onset of an instability at large values of $z$. First recall that
the mass flux at the boundary of the vacancy island in 
the normal direction $\vec{n}$ (pointing inwards) 
is the sum of the diffusive flux
$$
 j_d = - D\nabla c \cdot \vec{n}
$$
and the flux caused by electromigration
$$
 j_e = \frac{D}{\xi} c_\mathrm{eq} \vec{n} \cdot \hat{x},
$$
where, as usual, $\hat{x}$ denotes the unit vector in $x$-direction.
Next we consider the two simplified cases of (a) $j_e = 0$, i.e. no electric field
and (b) $j_d = 0$, i.e, no diffusion, to argue that
$j_d$ is stabilizing while $j_e$ is destabilizing:
\begin{itemize}
\item[(a)] No electric field: An outward bump of the boundary leads to a local
  maximum of the curvature and therefore a local minimum of the adatom density
  $c_\mathrm{eq}$ along the boundary.  Due to the maximum principle, this will
  also be a local minimum of $c$ inside the island, leading to a diffusive
  flux towards the boundary, i.e., the bump will be filled. Thus the diffusive
  flux is stabilizing.
\item[(b)] No diffusion: Consider an outward bump of the boundary at the
  trailing (back) island edge.  As in case (a), this leads to an increased
  curvature and therefore a decreased adatom density $c_\mathrm{eq}$ as
  compared to the front side.  The flux $j_e$ caused by electromigration is
  proportional to $c_\mathrm{eq}$ and therefore decreased at the back side,
  i.e. this part of the boundary is now drifting more slowly than the center
  of mass and the bump is growing.  Thus the electromigration flux
  destabilizes the trailing edge and stabilizes the leading edge of the
  island.
\end{itemize}
Finally consider the general case of an outward bump of the trailing boundary
in the presence of diffusion and electromigration.  Fixing all parameters,
except the strength of the electric field, $\xi^{-1}$, the diffusive flux
depends essentially on the geometry of the bump, whereas the decrease of $j_e$
is proportional to $\xi^{-1}$.  Thus we may expect that the bump will grow if
the electric field is large enough.

The argument also elucidates the role of the capillarity parameter $\Gamma$ in
this problem, which is mainly to translate variations of the boundary
curvature into variations of the adatom concentration in the interior of the
island.  The latter in turn underlies both the stabilizing effect of the
diffusive current and the destabilizing effect of electromigration. In
contrast to the instabilities in the PD regime, which can be described in
terms of a competition between capillarity and electromigration
\cite{Schimschak1998,Kuhn2005a,Kuhn2005b}, here capillarity plays a largely
neutral part, which explains (in hindsight) why the linear stability
properties are independent of $\Gamma$ (see Section \ref{s:linStability}).
%

%
%
\section{Conclusions}
\label{s:conclusions}

In this paper, we have presented a detailed study of a continuum model for the
electromigration-driven shape evolution of single-layer vacancy islands
mediated by internal terrace diffusion.  Significant shape deformations, such
as the elongation transverse to the field direction described in Section
\ref{s:steadyStates} or the pinch-off instability discussed in
Sect.\ref{s:nonLinEv}, were found to require dimensionless electromigration
forces $z = R_0/\xi$ significantly larger than unity. Unfortunately this
implies that these phenomena will be difficult to realize experimentally, at
least for the surfaces commonly used in this context. To give an example, the
maximal electromigration bias that can be achieved on the Cu(100) surface has
been estimated \cite{Mehl2000} to be on the order of $E_\mathrm{bias} \approx
10^{-5}$ eV for a diffusion hop between nearest neighbor sites, corresponding
to a characteristic length scale $\xi/a = k_{\mathrm{B}} T/E_\mathrm{bias}
\approx 2.5 \times 10^3$. An island with $z = 10$ would thus contain about $2
\times 10^9$ atoms, which is four orders of magnitude larger than the size at
which strong shape deformations and electromigration-induced oscillatory
dynamics have been predicted in the PD regime \cite{Rusanen2006}.

From the broader perspective of the theory of moving boundary value problems,
our results are of interest because they add another example to the list of
cases in which the standard tool of linear stability analysis fails to
correctly predict the stability properties of the full nonlinear dynamics. The
behavior of the interior model in the TD limit is similar in many respects to
the void migration problem in the PD regime which was studied in
\cite{Schimschak1998,Schimschak2000}. In both cases the circular solution is
linearly stable for arbitrary values of the electromigration force, but a
nonlinear instability occurs when the system is subjected to finite amplitude
perturbations, and the threshold for the nonlinear instability decreases with
increasing force.  Moreover, as in the case of void migration, a distorted
island either relaxes back to the circular shape or evolves towards pinch-off.
This suggests (as has been proven for voids \cite{Cummings2001}) that no
non-circular stationary states exist. However, in contrast to the void
migration problem, in the present study we found no evidence for transient
amplification of linear perturbations related to the non-normality of the
eigenvalue problem.

In this context it is worth mentioning the problem of two-dimensional
ionization fronts, which shares some of the features of both void and island
migration \cite{Meulenbroek2005}.  In this system a closed curve representing
the ionized region (a ``streamer'') evolves in response to a Laplacian
potential in the exterior domain. As in the void migration problem, a constant
potential gradient far away from the streamer represents the driving electric
field. However, the boundary condition at the front is similar to that
employed in the present work, (\ref{eq:fluxes},\ref{eq:bc},\ref{eq:vel})
with $c_{\mathrm{eq}} = \mathrm{const.}$, i.e. $\Gamma = 0$, and the kinetic
length corresponds to the width of the ionization front.  Again, circles
translating at constant velocity are solutions of the problem.  In the special
case where the kinetic length equals the radius of the streamer the linearized
dynamics can be solved exactly, and it is found that the circle is always
stable.  Standard linear stability analysis nevertheless fails, because smooth
initial perturbations do not decay exponentially \cite{Ebert2006}.  Further
exploration of the relationship between these three boundary value problems
seems like a promising direction for future research.


\appendix

\section{Bessel functions}
We summarize some properties of the
modified Bessel functions of imaginary argument $I_n$. For integer $n$ the
functions $I_n$ are symmetric with respect to the index
\begin{equation}
  \label{eq:besselSym}
   I_{-n}(r) = I_n(r).
\end{equation}
The derivative is
\begin{equation}
  \label{eq:besselRec}
   \tfrac{d}{dr} I_n(r) =
\tfrac{1}{2}(I_{n-1}(r) + I_{n+1}(r)).
\end{equation}
There is the recursion relation
\begin{equation}
  \label{eq:besselRec2}
   r I_{n-1}(r) - r I_{n+1}(r) =
 2n I_n(r),
\end{equation}
and the generating function is
\begin{equation}
  \label{eq:besselExp}
   \exp(r\cos\theta ) =
 \sum_{n=-\infty}^{\infty}  I_n(r)\exp(in\theta).
\end{equation}

\section{Normal velocity of an ellipsoidal island}
\label{a:Mathieu}
To calculate the concentration profile inside an ellipsoidal island, we
introduce elliptic coordinates
\begin{align*}
  x=\alpha \cosh(u) \cos(w), \\
  y=\alpha \sinh(u) \sin(w),
\end{align*}
where $u$ and $w$ are the radial and angular coordinates, respectively. The
line $u\equiv u_0 = \mathrm{const.}$ is an ellipse with aspect ratio
$\tanh(u_0)$ and the parameter $\alpha$ determines the size of this ellipse.
Curvature and normal vector are
\begin{align*}
  \kappa&=-\frac{1}{\alpha
    g^{\frac{3}{2}}}\cosh(u_0)\sinh(u_0),\\
  \vec{n}&=\frac{1}{\sqrt{g}}\left( \begin{array}{c}
      \sinh(u_0) \cos(w) \\
      \cosh(u_0) \sin(w)
\end{array} \right),\\
\end{align*}
with $g=\cosh^2(u_0)-\cos^2(w)$.  The Helmholtz equation now reads
\begin{align*}
  (\partial_u^2+\partial_w^2)f=\alpha^2(\sinh^2(u)+\sin^2(w))f
\end{align*}
and separation $f(u,w)=f_u(u) f_w(w)$ leads to
\begin{align*}
  f_w''-\alpha^2\sin^2(w)f_w &=\lambda f_w , \\
  -f_u''+\alpha^2\sinh^2(u)f_u &=\lambda f_u.
\end{align*}
The solutions of the two equations are related by $f_w(ix)=f_u(x)$. Requiring
periodicity of $f_w$ determines a discrete set of values for the separation
parameter $\lambda$. The corresponding solutions are the Mathieu functions of
the first kind $\mathrm{ce}_n$ and $\mathrm{se}_n$ \cite{Gradshteyn}. Since
the ellipse is symmetric with respect to the field direction we only need the
$\mathrm{ce}_n$ which are even functions of the angular variable $w$. The
general solution is then
\begin{align} 
\label{eq:ellipticsol}
  c(u,w)=\exp\left(\frac{\alpha}{2\xi}\cosh(u)\cos(w)\right)\sum_{n=0}^{\infty}b_n
  \mathrm{ce}_n(w) \mathrm{ce}_n(iu).
\end{align}
The coefficients $b_n$ are determined by the boundary condition
(\ref{eq:bcTD}). The Mathieu functions are orthogonal and normalized according
to $\int_0^{2\pi} \mathrm{ce}_n^2(x)\;dx=\pi$. Thus, the $b_n$ can be found
in a way similar to Fourier coefficients via the integrals
\begin{align*} 
  b_n&=\frac{c^0_\mathrm{eq}}{\pi \mathrm{ce}_n(iu_0)} \int_0^{2\pi}
  \mathrm{ce}_n(w)\exp\left(-\frac{\alpha}{2\xi}\cosh(u_0)\cos(w)\right)\times
  \notag \\
  & \times (1-\Gamma \kappa) \; dw.
\end{align*}
These integrals are solved numerically. 
Finally, using the general solution
(\ref{eq:ellipticsol}) to calculate the flux to the boundary, 
 the normal velocity $v$ of the island edge is obtained as 
\begin{align}
  \nonumber v&=\frac{a^2 D}{\sqrt{g}}\exp\left(\frac{\alpha}{2\xi}\cosh(u_0)
    \cos(w)\right)\sum b_n \mathrm{ce}_n(w) \times\\
\label{eq:velEllipt}
&\times \left(\frac{1}{\alpha}\mathrm{Im}(\mathrm{ce}_n'(iu_0))+\frac{1}{\xi}
  \sinh(u_0)\cos(w)\mathrm{ce}_n(iu_0)\right).
\end{align}

\begin{acknowledgments} We are grateful to Ute Ebert, 
Yan Fyodorov, Olivier Pierre-Louis, Vakhtang Putkaradze and Gerhard Wolf for instructive
discussions and helpful suggestions. This work was supported by DFG within project
KR 1123/1-2 and within SPP 1253.
\end{acknowledgments}

\bibliography{pre_elmig2d}

\end{document}